\documentclass[conference]{IEEEtran}
\IEEEoverridecommandlockouts
\usepackage[colorlinks,
            linkcolor=blue,
            anchorcolor=red,
            citecolor=red]{hyperref}
\newcommand{\T}{{\scriptscriptstyle\mathsf{T}}}
\renewcommand{\H}{{\scriptscriptstyle\mathsf{H}}}
\usepackage{booktabs}
\usepackage{cite}
\usepackage{amsmath,amssymb,amsfonts}
\usepackage{algorithmic}
\usepackage{graphicx}
\usepackage{textcomp}
\usepackage{xcolor}
\def\BibTeX{{\rm B\kern-.05em{\sc i\kern-.025em b}\kern-.08em
T\kern-.1667em\lower.7ex\hbox{E}\kern-.125emX}}

\usepackage{bm}
\usepackage{subfigure}
\usepackage[ruled]{algorithm2e}
\newtheorem{lemma}{Lemma}
\graphicspath{{Figures/}}
\newcommand\Nt{\ensuremath{ N_{\rm t} }}
\newcommand\Nr{\ensuremath{ N_{\rm r} }}
\newcommand\Kcl{\ensuremath{\mathcal{K}}}
\newcommand\ssb{\ensuremath{ \mathbf{s} }}
\newcommand\Cs{\ensuremath{{\mathbb{C}}}}
\newcommand\Wb{\ensuremath{ \mathbf{W} }}
\newcommand\wb{\ensuremath{ \mathbf{w} }}
\newcommand\xb{\ensuremath{ \mathbf{x} }}

\begin{document}
\setlength{\abovedisplayskip}{3.25pt}
\setlength{\belowdisplayskip}{3.25pt}
 
\title{Low-Complexity Cram\'er-Rao Lower Bound and Sum Rate Optimization in ISAC Systems
}

\author{\IEEEauthorblockN{Tianyu Fang\IEEEauthorrefmark{1}, Nhan Thanh Nguyen\IEEEauthorrefmark{1}, Markku Juntti\IEEEauthorrefmark{1}}
\IEEEauthorblockA{\IEEEauthorrefmark{1}Centre for Wireless Communications, University of Oulu, P.O.Box 4500, FI-90014, Finland}
\small Email: \{tianyu.fang, markku.juntti, nhan.nguyen\}@oulu.fi
\thanks{This work was supported in part by the Research Council of Finland through the 6G Flagship program (grant no. 346208) and the DIRECTION project (grant no. 354901).}
}

\maketitle

\begin{abstract}
While Cram\'er-Rao lower bound is an important metric in sensing functions in integrated sensing and communications (ISAC) designs, its optimization usually involves a computationally expensive solution such as semidefinite relaxation. In this paper, we aim to develop a low-complexity yet efficient algorithm for CRLB optimization. We focus on a beamforming design that maximizes the weighted sum between the communications sum rate and the sensing CRLB, subject to a transmit power constraint. Given the non-convexity of this problem, we propose a novel method that combines successive convex approximation (SCA) with a shifted generalized power iteration (SGPI) approach, termed SCA-SGPI. The SCA technique is utilized to approximate the non-convex objective function with convex surrogates, while the SGPI efficiently solves the resulting quadratic subproblems. Simulation results demonstrate that the proposed SCA-SGPI algorithm not only achieves superior tradeoff performance compared to existing method but also significantly reduces computational time, making it a promising solution for practical ISAC applications.
\end{abstract}


\section{Introduction}
Integrated sensing and communications (ISAC) is an emerging paradigm that seeks to unify the traditionally separate domains of wireless communications and sensing, and it has attracted considerable attention from both academia and industry~\cite{ma2020joint,zhang2021overview}. A substantial body of research has focused on joint beamforming optimization in multi-user multi-antenna ISAC systems~\cite{liu2018mu, liu2020joint, nguyen2023jointssp,nhan2023multiuser,choi2024joint,fang2024beamforming, zhang2023isac, wang2023optimizing,liu2021cramer, song2023intelligent, xiong2023fundamental}. Common performance metrics for the communications functionality include the achievable (sum) rate or signal-to-interference-plus-noise ratio (SINR). On the other hand, various metrics can be used for sensing, such as the beampattern~\cite{liu2018mu, liu2020joint, nguyen2023jointssp, nhan2023multiuser}, signal-to-clutter-plus-noise ratio (SCNR)~\cite{choi2024joint,fang2024beamforming, zhang2023isac, wang2023optimizing}, Cramér--Rao lower bound (CRLB) of the target position estimate~\cite{liu2021cramer, song2023intelligent,xiong2023fundamental,zhou2024cram}. 

Notably, the CRLB is recognized as an important metric that provides a fundamental theoretical limit on the accuracy of parameter estimation. Semi-definite relaxation (SDR) has emerged as a popular technique for handling the CRLBs. Liu \textit{et al.}~\cite{liu2021cramer} considered the CRLB minimization problem subject to communications SINR and transmit power constraint. However, there are inherent challenges in using SDR for  beamforming designs involving the sum rate (SR) metric for communications functionality. To address these challenges, Chen \textit{et al.}~\cite{chen2024transmitter} and Zhu \textit{et al.}~\cite{zhu2023integrated} combined SDR with techniques like successive convex approximation (SCA) or weighted minimum mean square error (WMMSE) to achieve a locally optimal solution. Fractional programming (FP) is another widely used approach for CRLB optimization \cite{zhu2023integrated,Guo2023FP-ADMM,Liu2024SNR, chen2024fast}. For instance, in \cite{Guo2023FP-ADMM}, auxiliary variables are introduced to transform the fractional CRLB into a polynomial form, enabling reformulating the original problem into a multi-variable optimization problem, which is then solved using the alternating direction method of multipliers (ADMM) framework. Additionally, a globally optimal beamforming design leveraging the branch and bound method was proposed in~\cite{wang2024globally} for maximizing the weighted sum of SR and CRLB. Although offering superior performance, the branch and bound approach in~\cite{wang2024globally} requires an excessively high complexity, which increases exponentially with the number of communications users.

Most of the aforementioned methods have various limitations, such as the requirement of standard optimization toolboxes to solve disciplined convex problems, high computational complexity, and the introduction of additional auxiliary variables. To overcome these challenges, we herein propose an efficient beamforming design method leveraging the SCA framework. Specifically, our focus is on solving the fundamental problem of maximizing the weighted sum of the SR and the CRLB in a multi-user, multi-antenna monostatic ISAC system under a transmit power constraint. The main contributions of this work are twofold: 1) We propose a unified framework using SCA to address the non-convexity arising from the objective functions of both the communications and sensing functionalities. 2) By exploiting the structure of the approximated problems, we develop an efficient power iteration-like algorithm to solve each approximated subproblem. Simulation results demonstrate that the proposed method achieves a superior trade-off performance compared to the benchmark, with significantly reduction in computational complexity.

\section{System Model and Problem Formulation}
\subsection{Signal Model}
We consider a downlink monostatic ISAC system, where a base station (BS) is equipped with $ \Nt $ transmit antennas and $ \Nr $ radar receive antennas, arranged as separate uniform linear arrays (ULAs) with half-wavelength spacing. A two-dimensional model is assumed, focusing on azimuth angle, under the premise that the environment and the signals of interest can be adequately represented in a plane. The BS simultaneously transmit data signals to $K$ single-antenna communications users, indexed by the set $\Kcl=\{1,\ldots,K\}$, while also utilizing these data signals to estimate a point-like sensing target over $L$ time slots, indexed by $\mathcal L=\{1,\ldots,L\}$. Let $ \ssb_{l}=[s_{1l},\ldots, s_{Kl}]^\T \in \Cs^{K\times 1} $ represent the transmitted symbol vector in the $l$-th time slot and $ \Wb=[\wb_1,\ldots,\wb_K]\in\Cs^{\Nt\times K} $ denote the corresponding beamforming matrix at the BS. The transmitted signal at the $l$-th time slot is then given by $ \xb_l= \Wb \ssb_l$, $\forall l\in\mathcal L$. By defining $\mathbf S=[\ssb_1,\ldots,\ssb_L]\in\mathbb C^{K\times L}$, the overall transmitted signal can be expressed as $\mathbf X=\mathbf W\mathbf S$.
It is assumed that the data streams are independent of each other, leading to $\mathbf S\mathbf S^\H\approx L\mathbf I$ \cite{liu2021cramer}.

\subsubsection{Communications Signal Model}
The signal received by communications user $ k $ at time slot $l$ is given by
\begin{equation}
     y_{kl}=\mathbf h_k^\H\mathbf w_ks_{kl}+\mathbf h^\H_k\sum_{j\neq k}^K\mathbf w_js_{jl} +n_{kl}, \forall k\in\mathcal K, \forall l\in\mathcal L,
\end{equation}
where $ \mathbf h_k\in\mathbb C^{\Nt\times 1} $ is the channel vector between the BS and user $ k $, and $ n_{kl}\sim\mathcal{CN}(0,\sigma_{\mathrm{c} k}^2)  $ is additive white Gaussian noise. The communications channels are assumed to be perfectly known at the BS and remain invariant throughout the entire transmission block. Consequently, the achievable rate in nats/s/Hz for user $ k $ is given by
\begin{equation}\label{rate}
	R_{ k}=\log\left(1+\frac{|\mathbf h_k^\H\mathbf w_k|^2}{\sum_{j\neq k}^K|\mathbf h^\H_k\mathbf w_j|^2+\sigma_{\mathrm{c} k}^2}\right),\forall k\in\mathcal K.
\end{equation} 
The SR of all the communications users, given as $\sum_{k=1}^K R_k$, is used as the metric to evaluate the system's overall communications performance.

\subsubsection{Sensing Signal Model}
The received echo signal at the BS is given by
\begin{equation}\label{sensing signal matrix}
    \mathbf Y_{\mathrm{s}}=\alpha \mathbf G(\mathbf \theta) \mathbf X+\mathbf N_{\mathrm{s}},
\end{equation}
where $\alpha$ refers to the radar cross-section, and $\mathbf N_{\mathrm{s}}\in\mathbb C^{\Nr\times L}$ represents the noise matrix at the receive antennas, with entries distributed as $\mathcal{CN}(0,\sigma_s^2)$. Here, $\mathbf G(\mathbf \theta)$ denotes the two-way channel in the desired sensing direction, modeled as
\begin{equation}\label{eq_sensing_channel}
      \mathbf G(\theta)=\mathbf a_{\mathrm{r}}(\theta)\mathbf a_{\mathrm{t}}^\H(\theta),
\end{equation}
where $ \mathbf a_{\mathrm{r}}(\cdot)$ and $\mathbf a_{\mathrm{t}}(\cdot) $ are the receive and transmit array steering vectors at the BS, and $\theta\in[-\pi/2,\pi/2]$ is the azimuth angle of the target relative to the BS. 

We use the CRLB associated with $\theta$ as the sensing performance metric, which is obtained from the inverse of the Fisher Information Matrix (FIM). To derive the FIM, we begin by vectorizing \eqref{sensing signal matrix} as
\begin{equation}
     \mathbf y_{\mathrm{s}}= \mathbf x_{\mathrm{s}}+\mathbf n_{\mathrm{s}},
\end{equation}
where $\mathbf y_{\mathrm{s}}=\mathrm{vec}(\mathbf Y_{\mathrm{s}})$, $\mathbf x_{\mathrm{s}}=\mathrm{vec}(\alpha\mathbf G\mathbf X)$ and $\mathbf n_{\mathrm{s}}=\mathrm{vec}(\mathbf N_{\mathrm{s}})$. Define the vector of the unknown parameters as $\bm\omega=[\theta,\widehat{\bm\alpha}]^\T\in\mathbb R^{3\times 1}$ from \eqref{sensing signal matrix}, where $\widehat\alpha=[\Re\{\alpha\},\Im\{\alpha\}]^\T$ stacks the real and imaginary parts of $\alpha$. The FIM for estimating all parameters in $\bm\omega$, denoted by $\mathbf F$, can be obtained by
\begin{equation}
    \mathbf F=2\Re \left\{\frac{\partial \mathbf x_{\mathrm{s}}^\H}{\partial \bm\omega}\mathbf R_{\mathrm s}^{-1}\frac{\partial \mathbf x_{\mathrm{s}}}{\partial \bm\omega} \right\}=\frac{2}{\sigma_{\mathrm s}^2}\Re \left\{\frac{\partial \mathbf x_{\mathrm{s}}^\H}{\partial \bm\omega}\frac{\partial \mathbf x_{\mathrm{s}}}{\partial \bm\omega} \right\},
\end{equation}
where $\mathbf R_{\mathrm{s}} =\sigma_{\mathrm s}^2 \mathbf I_{N_r L}$. Further, $\mathbf F $ can be partitioned into blocks as
\begin{align*}
		\mathbf F= 
		\begin{bmatrix}
			 F_{\theta\theta}   & \mathbf F_{\theta\widehat{\bm\alpha}}\\
			\mathbf F_{\theta\widehat{\bm\alpha}}^\T &\mathbf F_{\widehat{\bm\alpha}\widehat{\bm\alpha}}
		\end{bmatrix},
\end{align*}
with the (block) elements of $\mathbf F$ given by~\cite{bekkerman2006target}
\begin{equation*}
    \begin{aligned}
		F_{\theta\theta} &= \kappa |\alpha|^2 \mathrm{tr}(\dot{\mathbf G}_{\theta} \mathbf R_x \dot{\mathbf G}^\H_{\theta}) \in \mathbb R^{1 \times 1},  \\
		\mathbf F_{\theta \widehat{\bm\alpha}} &= \kappa \Re\{\alpha^* \mathrm{tr}(\mathbf G \mathbf R_x \dot{\mathbf G}^\H_{\theta})[1,1j] \} \in \mathbb R^{1 \times 2}, \\
		\mathbf F_{\widehat{\bm\alpha}\widehat{\bm\alpha}} &= \kappa \mathrm{tr}(\mathbf G \mathbf R_x \mathbf G^\H )\mathbf I  \in \mathbb R^{2 \times 2},  
	\end{aligned}
\end{equation*}	
where $\kappa \triangleq \frac{2L}{\sigma_s^2}$, $\mathbf R_x =\mathbf W\mathbf W^H$, and $$
		\dot{\mathbf G}_{\theta} = \dot {\mathbf a}_r(\theta) \mathbf a_t^\H(\theta) + \mathbf a_r(\theta) \dot{\mathbf a}_t^\H(\theta).$$

\subsection{Problem formulation}
Our objective is to optimize the transmit beamforming matrix $ \mathbf W $ to achieve good communications--sensing performance tradeoff. While using the SR to evaluate communications performance, we employ \textit{Trace-Opt} criterion and use $\mathrm{tr}(\mathbf{F}^{-1})$ as the sensing performance metric, following \cite{Li2008range}. By minimizing $\mathrm{tr}(\mathbf{F}^{-1})$ in the target optimization, we aim to reduce the CRLBs associated with all estimated parameters. It is worth noting that in~\cite{liu2021cramer}, the first entry of $\mathbf{F}^{-1}$ is directly minimized, as the focus is on estimating the azimuth angle of the target. However, as shown in~\cite{Li2008range}, considering all unknown parameters can lead to better sensing performance. As alternatives to the CRLB, the trace, determinant, and largest eigenvalue of the $\mathbf{F}^{-1}$ have been commonly considered in the literature \cite{chen2024transmitter,zhu2023integrated,Liu2024SNR, chen2024fast,wang2024globally}. Among these, minimizing the trace of $\mathbf{F}^{-1}$ offers the best overall performance in terms of accuracy and robustness \cite{Li2008range}. Therefore, we adopt this metric for evaluating sensing performance. 

The joint design problem can be formulated as:
 \begin{equation}\label{P1}
 \max_{ \mathbf W\in\mathcal S}\,\, \delta\sum_{k=1}^K R_k- \mathrm{tr}(\mathbf{F}^{-1}) 
 \end{equation}
where $ \mathcal S=\{\mathbf W\in\mathbb C^{\Nt\times K}:\mathrm{tr}(\mathbf W\mathbf W^\H)= P_{\mathrm{t}}\} $, with $P_{\mathrm{t}} $ representing the transmit power budget, and $ \delta$ is a weight that controls the tradeoff between communications and sensing performance. The problem in \eqref{P1} is inherently NP-hard due to the presence of multiple non-convex fractional SINRs. Solving \eqref{P1} to find the global optimum would involve exponential computational complexity, as indicated in~\cite{wang2024globally}. Therefore, our goal is to develop an efficient approach to obtain a locally optimal solution for \eqref{P1}.

\section{Beamforming Optimization via Successive Convex Approximation }
\label{sec: algorithm}
The proposed solution to \eqref{P1} combines the SCA and shifted generalized power iteration (SGPI) approaches. The former allows us to approximate the non-convex objective function in \eqref{P1} with a surrogate quadratic objective function, which is then solved effectively with the SGPI algorithm. We present the detailed solution next.

\subsection{The SCA Method}
We first introduce two lemmas that form the foundation for approximating the non-convex objective function in  \eqref{P1}.
\begin{lemma}\label{lemma:lin app com}
    Function $\log\left( 1+\frac{|z|^2}{d}\right)$ with $z\in\mathbb C$ and $d\in\mathbb R_{++}$ can be lower bounded by its first-order Taylor expansion as follows \cite{fang2023optimal}:
    \begin{equation}
        \begin{aligned}
            \log\left( 1+\frac{|z|^2}{d}\right )&\geq \log\left(1+\frac{|z_0|^2}{d_0}\right)+2\Re\{\frac{z_0^*}{d_0} z\}\\
            &-\frac{|z_0|^2}{d_0(d_0+|z_0|^2)}(|z|^2+d)-\frac{|z_0|^2}{d_0}.
        \end{aligned}
    \end{equation}
    The equality achieved at $(z,d)=(z_0,d_0)$, where  $(z_0,d_0)$ is a given feasible point.
\end{lemma}

\begin{lemma} \label{lemma:lin app sen}
    Function $\mathrm{tr}(\mathbf Z^{-1})$ with $\mathbf Z\in \mathbb S_{++}$ can be lower bounded by its first-order Taylor expansion as \cite{sun2017major}
    \begin{equation}
        \mathrm{tr}(\mathbf Z^{-1})\geq -\mathrm{tr}(\mathbf Z_0^{-1}\mathbf Z\mathbf Z_0^{-1} )+\mathrm{tr}(\mathbf Z_0^{-1}),
    \end{equation}
    with the equality achieved at $\mathbf Z=\mathbf Z_0$, where $\mathbf Z_0$ is a given feasible point.
\end{lemma}

By invoking Lemma \ref{lemma:lin app com} to $R_k$ given in \eqref{rate}, with $z=\mathbf h_k^\H\mathbf w_k$ and $d=\sum_{j=1,j\neq k}^K|\mathbf h_k^\H\mathbf w_j|^2+\sigma_{\mathrm c}^2$, we construct a surrogate function at iteration $t$ as follows
\begin{align}\label{surrogate: com}
    f_k^{[t]}&=\log(1+\xi_{k}^{{[t]}})+2\Re\{\mathbf h_k^\H\mathbf w_k\eta_{ k}^{{[t]}} \} \nonumber \\
		&\hspace{2cm}-\beta_k^{{[t]}}\left(\sum_{j=1}^{K}|\mathbf h_k^\H\mathbf w_j|^2+\sigma_{\mathrm{c} k}^2  \right)-\xi_{k}^{[t]},
\end{align}
where $\xi_k^{{[t]}}$, $\eta_k^{{[t]}}$, and $\beta_k^{{[t]}}$ are auxiliary variables, given as
\begin{subequations}\label{update auxilary communications}
    \begin{align}
        \xi_k^{{[t]}}&=\frac{|\mathbf h_k^\H\mathbf w_k^{[t]}|^2}{\sum_{j=1,j\neq k}^K|\mathbf h_k^\H\mathbf w_j^{[t]}|^2+\sigma_{\mathrm{c}}^2},\\
        \eta_k^{{[t]}}&=\frac{\xi_k^{[t]}}{\mathbf h_k^\H\mathbf w_k^{[t]}},\beta_k^{{[t]}}=\frac{\xi_k^{{[t]}}}{\sum_{j=1}^K|\mathbf h_k^\H\mathbf w_j^{[t]}|^2+\sigma_{\mathrm c}^2}.
    \end{align}
\end{subequations}
Similarly, by applying Lemma \ref{lemma:lin app sen} to $\mathrm{tr}(\mathbf{F}^{-1})$, we construct a surrogate function at iteration $t$ as $-\mathrm{tr}(\mathbf F \mathbf \Phi^{[t]})+\mathrm{tr}(\mathbf F^{[t]^{-1}})$,
where $\bm\Phi^{[t]}={\mathbf F^{[t]}}^{-2}$. Using these surrogate functions, problem \eqref{P1} can be approximated at a feasible point $\mathbf W^{[t]}$ as
 \begin{equation}
 	\label{P1:SCA}
 	\max_{ \mathbf W\in\mathcal S}\,\, \delta\sum_{k=1}^K f_k^{[t]}+ \mathrm{tr}(\mathbf F\bm\Phi^{[t]})-\mathrm{tr}(\mathbf F^{[t]^{-1}}). 
 \end{equation}
Although \eqref{P1:SCA} is still non-convex, it is now a quadratic problem with respect to $\mathbf W$. This form can be handled efficiently via the SGPI method as elaborated next.

\subsection{The SGPI Method}
\label{sec: complexity}
With some algebraic manipulation, \eqref{P1:SCA} can be expressed as a standard quadratic problem as follows:
\begin{align}\label{SubW}
	\max_{ \mathbf W\in\mathcal S}\,\,\,\,\mathrm{tr}(\mathbf W\mathbf W^\H\mathbf A)+2\delta\Re\{\mathrm{tr}(\mathbf W \bm\Sigma_1^\H\mathbf H^\H )\},
\end{align}
where
$\mathbf\Sigma_1=\mathrm{diag}\{\eta_1,\ldots,\eta_K \}$, $\mathbf H=[\mathbf h_1,\ldots,\mathbf h_K ]$, $\mathbf A=\lambda\mathbf I+\frac{1}{2}(\mathbf Q+\mathbf Q^\H)-\delta\mathbf H\bm\Sigma_2\mathbf H^\H$, $\mathbf \Sigma_2=\mathrm{diag}\{\beta_1,\ldots,\beta_K \}$, 
and 
\begin{align*}
\mathbf Q&=\kappa\big(\phi_{11}|\alpha|^2\dot{\mathbf G}^\H\dot{\mathbf G}+2(\phi_{12}+1j\cdot\phi_{13})\alpha^*\dot{\mathbf G}^\H\mathbf G \nonumber \\
&\hspace{4cm} +(\phi_{22}+\phi_{33})\mathbf G^\H\mathbf G\big).
\end{align*}
Here, $\phi_{ij}$ representing the $(i,j)$-th entry of the auxiliary matrix $\bm\Phi$, and $\lambda$ is a shift parameter chosen to ensure that $\mathbf A$ is a positive semi-definite matrix. Note that the superscript for auxiliary variable $\xi_k,\eta_k,\beta_k$ and $\mathbf \Phi$ have been omitted for brevity. Additionally, $\lambda$ can be any arbitrary constant greater than or equal to the dominant eigenvalue of $\delta\mathbf H\bm\Sigma_2\mathbf H^\H-\frac{1}{2}(\mathbf Q+\mathbf Q^\H)$~\cite{golub2013matrix}. To solve \eqref{SubW}, we introduce the following lemma.
\begin{lemma}\label{SCA}
For any given positive semi-definite Hermitian matrix $ \mathbf B \in\mathbb{C}^{N_{\mathrm{t}}\times N_{\mathrm{t}}}$ and matrix $\mathbf W$, the function $\mathrm{tr}(\mathbf W\mathbf W^\H\mathbf B)$ can be lower bounded by it's first-order Taylor expansion as
\begin{equation}\label{key}
\mathrm{tr}(\mathbf W\mathbf W^\H\mathbf B)\geq 2\Re\{\mathrm{tr}(\mathbf W_0\mathbf W^\H\mathbf B) \}-\mathrm{tr}(\mathbf W_0\mathbf W_0^\H\mathbf B ),
\end{equation}
where $ \mathbf W_0$ is a given feasible point, and the equality is achieved if and only if $ \mathbf W=\mathbf W_0$.
\end{lemma}

Based on Lemma \ref{SCA}, we can derive a linear approximation of problem \eqref{SubW} at iteration $n$ for the inner-layer SGPI method as
\begin{equation}\label{linear approximation}
	\max_{\mathbf W\in\mathcal S} \Re\{\mathrm{tr}(\mathbf W\bm\Sigma_1^\H\mathbf H^\H +\mathbf W^\H\mathbf A\mathbf W^{[n]}) \},
\end{equation}
The optimal solution of problem \eqref{linear approximation} is then given by
\begin{equation}\label{Update W}
	\mathbf W^{[n+1]}= \bm\Pi_{\mathcal S}\left(\mathbf H\mathbf \Sigma_1+\mathbf A \mathbf W^{[n]} \right),
\end{equation}
where $ \bm\Pi_{\mathcal S}(\cdot) $ denotes the projection of given point $ \mathbf W$ onto set $ \mathcal S $, i.e., $\bm\Pi_{\mathcal S}({\mathbf P})\triangleq \sqrt{P_{\mathrm{t}}/\mathrm{tr}({\mathbf P} {\mathbf P}^\H)}{\mathbf P}$. For clarity, we summarize the proposed SCA-SGPI method for solving problem \eqref{P1} in Algorithm \ref{al1}. The algorithm begins with an initial non-zero feasible matrix $\mathbf W^{[0]}$. In each iteration, the auxiliary variables $\xi_k,\eta_k,\beta_k, \forall k \in\mathcal K$ and $\mathbf\Phi$ are updated based on \eqref{update auxilary communications}, followed by the update of the transmit beamforming matrix $\mathbf W$. This process continues iteratively until the objective value in \eqref{P1} converges. We skip the detailed convergence analysis of Algorithm \ref{al1} due to limited space. However, numerical results will be shown to verify the convergence. 
\setlength{\textfloatsep}{7pt}	
\begin{algorithm}[t!]
    \small
	\textbf{Initialize}: $t\leftarrow0$, $\mathbf{W}^{[0]}$\;
	\Repeat{The objective value in \eqref{P1} converges.}{
		$t\leftarrow t+1$\;
		Update $ \xi_k^{[t]},\eta_k^{[t]} $ and $ \beta_k^{[t]}$ according to \eqref{update auxilary communications} for $\forall k\in\mathcal K$\;
		Update $\mathbf \Phi^{[t]}={\mathbf F^{[t]}}^{-2}$\;	
	\For{$n=1:I_2$}{
 Update $\mathbf W^{[t]} $ by \eqref{Update W}\;
 }	
	}	
	\caption{The proposed SCA-SGPI Algorithm}
	\label{al1}				
\end{algorithm}
\subsection{Complexity Analysis}

The complexity for each iteration of Algorithm \ref{al1} is dominated by the update of the beamforming matrix, which is $\mathcal{O}(I_2 K N_t^2)$. Here, $I_2$ denotes the number of iterations in the inner loop. Therefore, the overall complexity of the proposed SCA-SGPI algorithm is $\mathcal{O}(I_1 I_2 K N_t^2)$, with $I_1$ representing the number of iterations in the outer loop. It is observed that Algorithm \ref{al1} has significantly lower complexity than the widely employed SDR techniques in the literature with a complexity of $\mathcal O(I_1K^{3.5}\Nt^{6.5})$ \cite{wang2022par}.

\section{Numerical Results}
\label{sec: numerical results}
In the numerical examples, we set $\Nt=16$, $\Nr=20$, $K=4$, $L=30$ \cite{liu2021cramer}. The transmit power is set to $P_t=10$ dBm and the noise variances are $\sigma_s^2=\sigma_{ck}^2=0$ dBm. We adopt the Rayleigh fading model for the communications channel. Furthermore, the steering vector $\mathbf a_t(\theta)$ in \eqref{eq_sensing_channel} is modeled as $\mathbf a_t(\theta)=[e^{-1j\frac{\Nt-1}{2}\pi\sin{\theta}},e^{-1j\frac{\Nt-3}{2}\pi\sin{\theta}}\ldots,e^{1j\frac{\Nt-1}{2}\pi\sin{\theta}} ]^\T$;  
$\mathbf a_r(\theta)$ is modeled similarly\cite{liu2021cramer}. For the sensing component, the reflection coefficient is set to $\alpha=0.1 \left(\sqrt{\frac{2}{3}}+1j\sqrt{\frac{1}{3}}\right)$, and the target angle is $\theta=0$. In Algorithm \ref{al1}, $I_2=20$ inner iterations are performed, and the convergence tolerance for the outer loop is set to $10^{-4}$. All the presented results are averaged over 100 channel realizations.
\begin{figure}[t]
\small
    \centering
    \hspace{-2mm}
    \subfigure[Outer loop.]
    {\label{fig:convergence_outer}\includegraphics[width=0.25\textwidth]{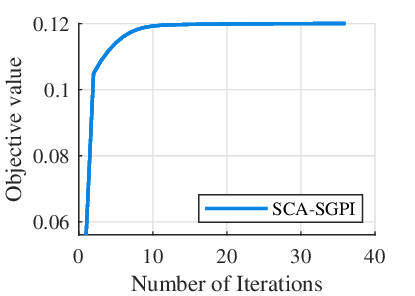}}
     \hspace{-5mm} 
    \subfigure[Inner loop.]
    {\label{fig:convergence_inner} \includegraphics[width=0.25\textwidth]{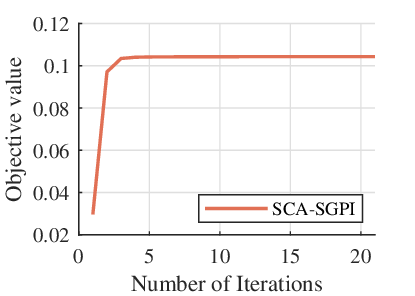}}
    \vspace{-0.25cm}
    \caption{Convergence of the inner and outer loops of Algorithm \ref{al1} with $\Nt=16$, $\Nr=20$, $K=4$, and $L=30$.}
    \label{fig:SE performance} 
        \vspace{-0.15cm}
\end{figure}

In Fig.\ \ref{fig:SE performance}, we show the convergence behavior of both the outer and inner loops in Algorithm \ref{al1}, with $\delta=0.01$. It is observed that the the objective value monotonically increases and converges after only $5$ iterations (for the inner loop) and $10$ iterations (for the outer loop). In the following figures, we use the SR and CRLB for estimating the angle $\theta$, i.e., the first diagonal elements of $\mathbf F^{-1}$, to show the communications and sensing performance, respectively.

\begin{figure}[t]
\vspace{-0.2cm}
\center
\includegraphics[width=0.45\textwidth]{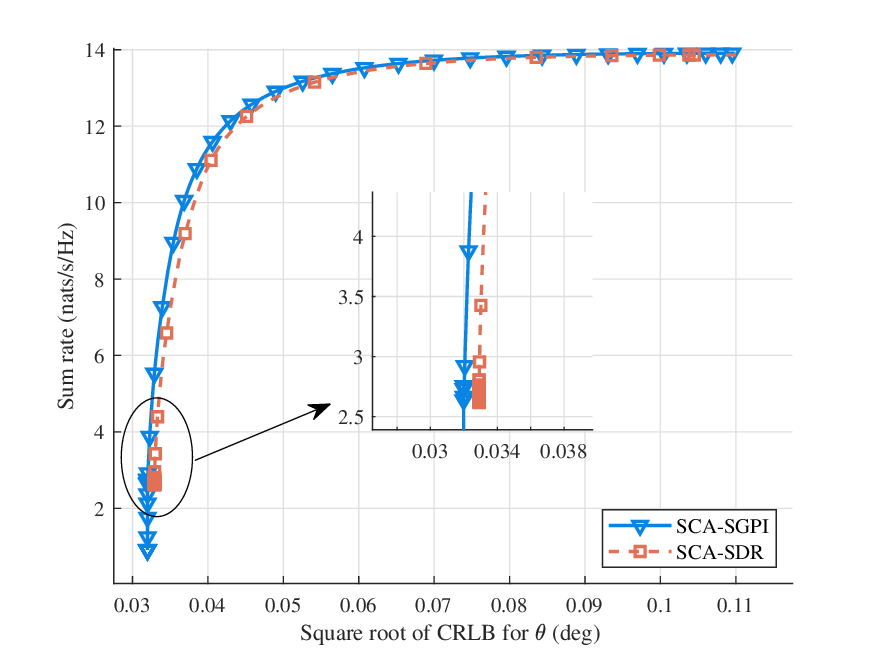}
     \hspace{-5mm} 
    \vspace{-0.25cm}
    \caption{Tradeoff region between communications and sensing with $\Nt=16$, $\Nr=20$, $K=4$, and $L=30$.}
    \label{fig:tradeoff_region}
\end{figure}

In Fig.\ \ref{fig:tradeoff_region}, we compare the communications--sensing tradeoff regions of proposed SCA-SGPI scheme and the SCA-SDR method proposed in~\cite{chen2024transmitter}. The communications and sensing performances are obtained for $\delta \in [10^{-7}, 10^1]$.  
It is observed that the proposed SCA-SGPI scheme achieves slightly better performance compared to its SCA-SDR counterpart. Moreover, the SCA-SGPI scheme has a runtime of just \textbf{0.0362} seconds, significantly faster than the SCA-SDR method, which takes \textbf{9.0803} seconds.

\begin{figure}[t]
\vspace{-0.5cm}
\center
\includegraphics[width=0.45\textwidth]{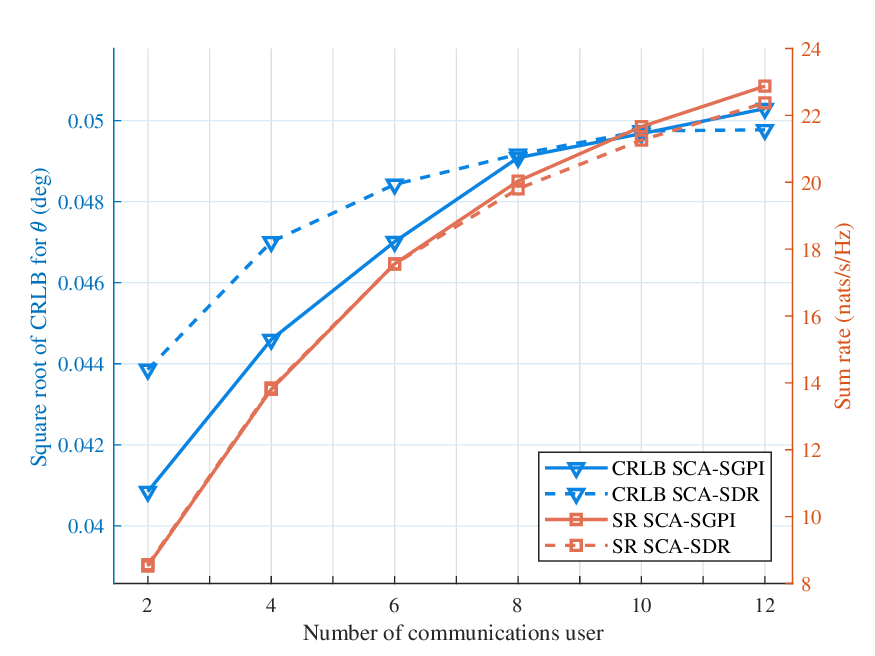}
     \hspace{-5mm} 
    \caption{Performance versus number of communications users.}
    \label{fig:Kuser}
\end{figure}

Fig.\ \ref{fig:Kuser} shows the communications and sensing performances for $K \in \{2,4,\ldots,12\}$ and $\delta=0.1$. It is shown that compared to the SCA-SDR counterpart, the proposed SCA-SGPI method achieves slightly better communications performance for large $K$ and much better sensing performance for small $K$. As $K$ increases, the communications sum rate improves significantly. However, when the sum rate becomes sufficiently large, it dominates the objective function in the ISAC beamforming design problem. This leads to degraded sensing accuracy, as indicated by the increasing value of $\mathrm{tr}(\mathbf{F}^{-1})$. 

In Table\ \ref{tab:Kuser}, we present the   run time required for obtaining the results in Fig\ \ref{fig:Kuser}. It is clear that the proposed algorithm performs significantly faster than the SCA-SDR approach, and the run time reduction becomes more significant as $K$ increases. This observation aligns well with the comparison of the two schemes in terms of computational complexity. Specifically, the complexity of the SCA-SDR algorithm scales with $\mathcal O(K^{3.5})$, while that of the proposed SCA-SGPI scales with $\mathcal{O}(K)$. These conclude the computational efficiency of the proposed method.

  \begin{table}[t!]
		\begin{center}
			\caption{Run time (seconds) of the SCA-SGPI and SCA-SDR algorithms for $K \in \{2,4,\ldots,12 \}$ and $\delta=10^{-1}$.}\vspace{-0.35cm}
			\label{tab:Kuser}
			\resizebox{\linewidth}{!}{\begin{tabular}{ccccccc}
				\toprule
				Algorithms & $K = 2$&$K = 4$&$K = 6$&$K = 8$&$K = 10$&$K = 12$
      \\
				\hline\hline
				\textbf{SCA-SGPI}& \textbf{0.0270}&\textbf{0.0185}&\textbf{0.0137}&\textbf{0.0117}&\textbf{0.0134}&\textbf{0.0143} \\
    \hline
    		SCA-SDR & 6.022&10.45&19.97&73.13&131.18&197.07 \\
				\bottomrule
			\end{tabular}}
		\end{center}
	\end{table}

\section{Conclusion}

We have presented an efficient ISAC beamforming design based on the SCA and SGPI approaches to overcome the high computational complexity introduced by existing SDR techniques. By leveraging efficient lower bound as surrogates of the objective function, we propose an low-complexity numerical algorithm for maximizing the weighted sum of SR and the trace of the inverse of the FIM. Simulation results shows the fast convergence of the proposed SCA-SPGI method. In particular, it achieves better communications--sensing performance tradeoff with substantially lower complexity and faster excution compared with the conventional SCA-SDR method. The proposed algorithm can be readily generalized to multi-target ISAC systems, which will be the focus of our future research.




\newpage
	\bibliographystyle{IEEEbib} 
	\bibliography{IEEEabrv,reference}

\end{document}